\documentclass[aps,prl,twocolumn,showpacs]{revtex4-2}
\usepackage{amssymb}
\usepackage{amsmath}
\usepackage{graphicx}
\usepackage{color}

\begin{document}
\title{Photon statistics of resonantly driven spectrally diffusive quantum emitters}
\author{Aymeric Delteil, St\'ephanie Buil and Jean-Pierre Hermier}

\affiliation{ Universit\'e Paris-Saclay, UVSQ, CNRS,  GEMaC, 78000, Versailles, France.\\ {\color{white}--------------------} aymeric.delteil@universite-paris-saclay.fr{\color{white}--------------------}  }


\begin{abstract}
In the solid state, a large variety of single-photon emitters present high quality photophysical properties together with a potential for integration. However, in many cases, the host matrix induces fluctuations of the emission wavelength in time, limiting the potential applications based on indistinguishable photons. A deep understanding of the underlying spectral diffusion processes is therefore of high importance for improving the stability of the light emission. Here, we theoretically investigate the photon statistics of an emitter driven by a resonant laser, and subject to either of two qualitatively different stationary spectral diffusion processes -- a continuous diffusion process and a process based on discrete spectral jumps, both of which being known to model the spectral diffusion of various solid-state emitters. We show that the statistics of light emission carries several experimentally accessible signatures that allow to discriminate between the two classes of models, both at short times in the intensity correlation function, and at long times in the fluctuations of the integrated intensity. These results establish that resonant excitation combined with photon statistics offers a rich access to the spectral diffusion processes, yielding information that goes beyond the bare characterization of the inhomogeneous shape and noise correlation time. Incidentally, our findings shed a new light on recent experimental results of spectral diffusion of B centers in hexagonal boron nitride, providing more insight in their spectral diffusion mechanisms.

\end{abstract}

 \maketitle
\section{I. Introduction}

Solid-state single-photon sources, such as self-assembled quantum dots and color centers in wide gap crystals, are widely seen as major actors in the emerging quantum technologies~\cite{Aharonovich16, Pelucchi22, Couteau23}. They can act like individual atoms trapped in a solid matrix, which enables integration into devices. In turn, these quantum emitters are also sensitive to the solid-state environment, which can manifest itself in decoherence and spectral diffusion (SD) of the emitter transition. The investigation of SD processes is therefore crucial for understanding and improving the performance of solid-state single-photon emitters for applications. In particular, fluctuations of the emitter wavelength limit the number of indistinguishable photons emitted by an artificial atom.

Spectral diffusion is associated with a stochastic process where the random variable $\hbar \omega (t)$ describes the spectral position of the center energy. This varying emission line can be characterized using a variety of experimental techniques, depending on the amplitude and time scale of the fluctuations. When the characteristic time of SD is faster than the inverse count rate, spectroscopy falls short, such that more complex techniques have to be envisaged. Amongst the possible strategies, photon correlation Fourier spectroscopy~\cite{Brokmann06, Coolen07,Beyler13} and sub-linewidth filtering~\cite{Sallen10, Abbarchi12} can be used, albeit with intrinsic limitations in either temporal or spectral resolution. { Delay-dependent Hong-Ou-Mandel interference can also be performed~\cite{Thoma16}, but is limited to SPEs emitting close-to-indistinguishable photons, and require to integrate coincidences separately for each time delay, which in many cases can turn out experimentally impractical.}

Recently, we have experimentally demonstrated that a combination of resonant laser excitation and photon correlations allows to establish the presence of spectral diffusion as well as some of its characteristics~\cite{Fournier23PRB}. Indeed, this technique simultaneously provides very high spectral and time resolutions. The resonant laser drive converts spectral fluctuations into intensity fluctuations, yielding photon bunching that can be measured through the intensity autocorrelation function $g^{(2)}(\tau)$. { It offers the advantage to probe more than 10~orders of magnitude of time scales in a single take, and does not need indistinguishable photons, nor is it limited to zero-phonon-line emission.} Such experimental work would benefit from a detailed theoretical investigation of the statistics of the emitted light and its connection with the microscopic process causing SD. Photon correlations are indeed well known as a powerful tool to reveal rich signatures of a large variety of fundamental microscopic phenomena involving quantum emitters, such as single photon emission~\cite{Michler00}, indistinguishability~\cite{Santori02}, entanglement~\cite{Akopian06}, superradiance~\cite{Jahnke16}, blinking~\cite{Schroeder21} and phase transitions~\cite{Fink17}. 

In the present work, we investigate the relation between the spectral diffusion process of a quantum emitter and the associated photon statistics under resonant excitation. We consider two emblematic stationary Markovian processes that yield Gaussian inhomogeneous broadening: the Ornstein-Uhlenbeck process and the Gaussian random jump process. While the former depicts a continuous spectral diffusion of some solid-state emitters, the latter models emitters undergoing discrete spectral jumps. Section~II describes the theoretical framework of these two spectral diffusion mechanisms. Section~III is devoted to the analysis of relevant statistical properties of resonantly scattered light when the emitter is subject to either stochastic process. We identify several experimentally measurable signatures that allow to identify and characterize the diffusion process. Finally, in Section~IV we discuss the obtained results and we provide some insights into the spectral diffusion process of B centers in hBN~\cite{Fournier21, Gale22, Shevitski19}. The framework and methods developed in the present work can be straightforwardly extended to other SD models and other experimental protocols, as we also discuss in section~IV { and in the Supplemental Material}.

%

\section{II. Models for spectral diffusion processes}
\label{model&simu}

In the solid state, the frequency fluctuations of an emitter can have various origins depending on the nature of the emitter, the properties of the host material and its dimensionality. When it comes to processes yielding a static Gaussian inhomogeneous distribution, two main classes of noise are commonly identified, depending on whether the variations of the transition energy are continuous in time or experiences discrete spectral jumps at random times. These two models are described in the following.

\subsection{A. Continuous spectral diffusion: the Ornstein-Uhlenbeck process}

\begin{figure}[h]
  \centering
  \includegraphics[width=0.7\linewidth]{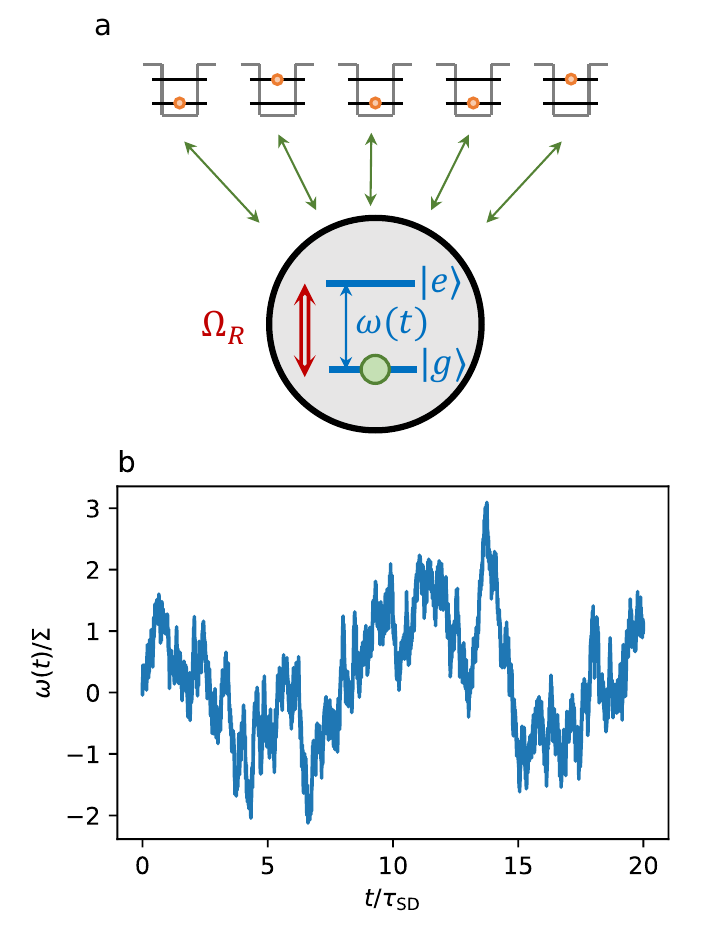}\\
  \caption{(a) Sketch of an emitter coupled to a fluctuating environment modeled by an ensemble of identical two-level systems, yielding a OU diffusion process. (b) Numerically generated spectral trajectory.}\label{fig1}
\end{figure}

In a wide range of physical systems, continuous diffusion can be well described by a Ornstein-Uhlenbeck (OU) process. The OU process is a Markovian and stationary Gaussian process, and as such, is completely characterized by its correlation function $\langle \omega(t + \tau)  \omega(t) \rangle = \Sigma^2 e^{-\tau/\tau_\mathrm{SD}}$, where $\Sigma$ is the standard deviation of the associated Gaussian probability distribution function. Originally derived in the frame of nuclear magnetic resonance by Kubo and Anderson~\cite{Kubo54, Anderson54}, it has been shown to also describe both charge noise and spin noise in self-assembled quantum dots~\cite{Berthelot06, Berthelot10, Kuhlmann13} as well as spectral diffusion in other condensed matter systems, such as ions or molecules embedded in solid-state matrices~\cite{Reilly93, DeVoe81}. This process emerges when the light source is coupled to a large ensemble of identical and independent two-level fluctuators~(Fig.~\ref{fig1}a), and is equivalent to the Wiener model of Brownian motion at short times.

Fig.~\ref{fig1}b shows an energy trajectory generated numerically using the stochastic differential equation $d\omega = dt (\omega_0 - \omega) / \tau_\mathrm{SD} + \Sigma dW_t$, where $\hbar \omega_0$ is the center energy of the inhomogeneous distribution, $\tau_\mathrm{SD}$ the spectral diffusion correlation time, and $W_t$ is the Wiener process {(the continuous-time stochastic process that describes standard Brownian motion~\cite{Durett})}. The numerical integration is performed based on the Euler-Maruyama method~\cite{Kloeden92}. An example of sample path is shown Fig~\ref{fig1}b. The trajectories exhibit a continuous drift such that, after a time larger than the correlation time $\tau_\mathrm{SD}$, the energy position is randomly found within a Gaussian envelope of standard deviation $\Sigma$.

\subsection{B. Gaussian random jump model}

\begin{figure}[t]
  \centering
  \includegraphics[width=0.7\linewidth]{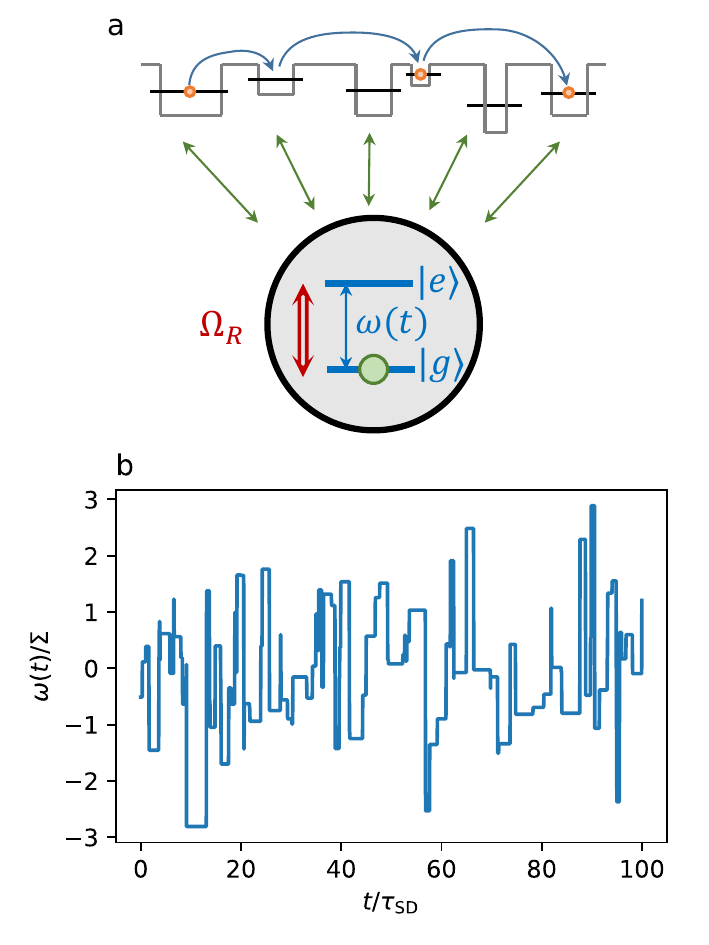}\\
  \caption{(a) Sketch of an emitter coupled to a fluctuating environment modeled by a charge hopping between various trap states. (b) Numerically generated spectral trajectory .}\label{fig2}
\end{figure}

Spectral diffusion can also take the form of discrete jumps, occuring at random times governed by a Poisson process, and sampling a Gaussian distribution of standard deviation $\Sigma$. Such stochastic process is termed Gaussian random jump (GRJ) model by Spokoyny \textit{et al.}~\cite{Spokoyny20} -- we also adopt this terminology in the present work. This process is also a Markovian and stationary process with correlation function $\langle \delta \omega(t) \delta \omega(0) \rangle = \Sigma^2 e^{-t/\tau_\mathrm{SD}}$, and therefore cannot be distinguished from the OU model on the sole basis of its probability density function (PDF) and its correlation function. It has been used to describe spectral fluctuations of nitrogen-vacancy centers in nanodiamonds~\cite{Wolters13}, molecules in crystalline matrices~\cite{Ambrose91}, color centers in hBN~\cite{Spokoyny20}, nitride quantum dots~\cite{Gao19}, perovskite~\cite{Utzat19} and CdSe nanocrystals~\cite{Beyler13} -- although, in the latter case, it is the spectral shifts (and not the spectral positions) that sample a Gaussian distribution -- which makes little difference at short times, but yields an additional long-time diffusion. Discrete jumps in spectral diffusion typically emerges when an emitter is coupled to a low number of carriers migrating in a manifold of heterogeneous trap states (Fig.~\ref{fig2}a), or a low number of neighboring systems whose configuration alternates within a continuum of states, such as ligand configurations in nanocrystals~\cite{Beyler13}.

The generation of a spectral trajectory is conceptually simple. The emitter energy $\hbar \omega(t)$ has a constant probability $dt/\tau_\mathrm{SD}$ to undergo a spectral jump during $dt$. When a jump occurs, the new frequency position is randomly drawn based on a Gaussian distribution of center energy $\hbar \omega_0$ and standard deviation $\Sigma$. An exemple of numerically generated trajectory is shown Fig.~\ref{fig2}b, where the discrete character of the spectral fluctuations is clear: $\omega(t)$ is piecewise constant and changes values at intervals of order $\tau_\mathrm{SD}$.

In the following section, we investigate the impact of these microscopic SD processes on the photon correlations of resonantly driven two-level emitters.

\section{III. Photon statistics of resonance fluorescence}
\label{RF}

In the absence of spectral diffusion, the population of a two-level atom driven by a laser close to resonance reaches a steady-state given by~\cite{Loudon}

\begin{equation}
\rho_{ee} =  \dfrac{1}{2} \dfrac{\Omega_R^2  T_1/T_2}{(\omega-\omega_L)^2 + T_2^{-2} + \Omega_R^2 T_1/T_2}
 \label{spectrum}
\end{equation}
where $\Omega_R$ is the laser Rabi frequency, $\hbar \omega_L$ its position in energy, $T_1$ the emitter lifetime, $T_2$ its coherence time and $\omega$ its center frequency.

The rate of scattered photons in the steady-state $C_{ss}(\omega)$ is proportional to $\rho_{ee}$, such that the intensity response to the laser excitation is a Lorentzian of linewidth $\Delta\omega_\mathrm{hom} = 2\sqrt{1 + \Omega_R^2T_1T_2}/T_2$. Conveniently, the laser power provides an external knob to tune the linewidth of the homogeneous lineshape through $\Omega_R$, a phenomenon known as power broadening.

{ In the following, unless explicitly stated, we suppose $\tau_\mathrm{SD} \gg T_1, 1/\Sigma$. This corresponds to cases where the spectral fluctuations are slower than both the emitter lifetime and the inverse inhomogeneous linewidth. Most experimental studies of solid-state emitters are consistent with these conditions, at the exception of some recent work demonstrating the achievement of fast SD that yields phenomena such as motional narrowing~\cite{Berthelot06, Berthelot09, Pont21, Bogaczewicz23}. Out of these specific regimes,} a continuously driven emitter is supposed to reach its steady state faster than the spectral fluctuations occur. This justifies the adiabatic approximation, in which the time-dependent intensity writes $C(t) = C_{ss}(\omega(t))$.

\begin{figure}[h]
  \centering
  \includegraphics[width=0.7\linewidth]{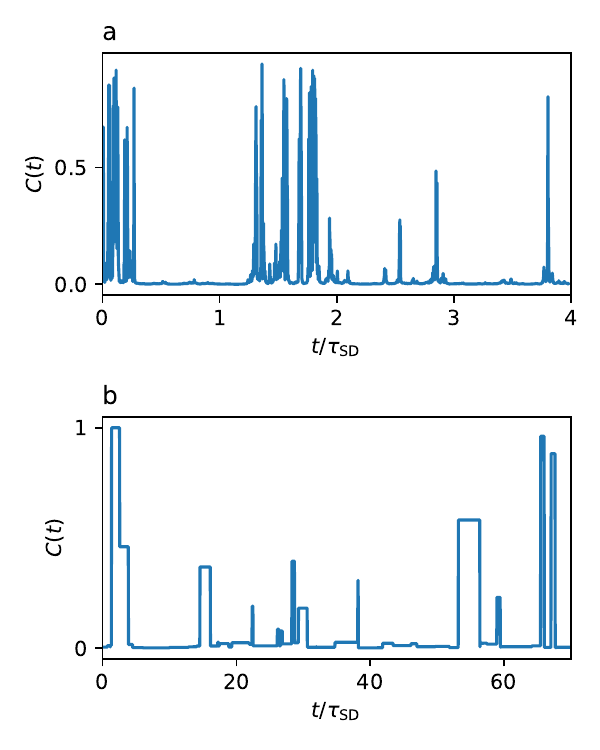}\\
  \caption{(a) Intensity time trace generated with the OU model. (b) Intensity time trace generated with the GRJ model.}\label{fig3}
\end{figure}

Fig.~\ref{fig3} shows two examples of intensity time traces generated with both a continuous and a jump process. In both cases, the signal exhibits an alternation of bright periods (when $\omega(t)\approx \omega_L$) and dark periods, albeit with a qualitatively different time dependence. Experimentally, these time-intensity curves are impossible to measure as long as the count rate is smaller than the inverse timescale of the intensity flucuations, which is the case in many practical situations. This justifies a general approach based on photon statistics.

\subsection{A. Intensity autocorrelation}

In this section, we focus on the intensity autocorrelation function $g^{(2)}(\tau)$, defined as 

%

\begin{equation}
g^{(2)}(\tau)  = \dfrac{\langle C(t) C(t + \tau) \rangle_t}{\langle C(t)  \rangle_t^2} = \dfrac{\langle C_{ss}[\omega(t)]C_{ss}[\omega(t + \tau)] \rangle_t}{\langle C_{ss}[\omega(t)]  \rangle_t^2}
 \label{g2}
\end{equation}
where $\langle ... \rangle_t$ denotes the time averaging, and where quantum effects at $\tau \lesssim T_1$ are neglected. 
As evidenced in prior work~\cite{Fournier23PRB}, resonantly driving a spectrally diffusive emitter leads to bunching, owing to the fact that detection of a photon at time $t_1$ informs on a close proximity between the emitter center frequency $\omega(t_1)$ and the laser frequency $\omega_L$. Therefore, the photon detection rate at time $t_2 \approx t_1$ is enhanced with respect to an uncorrelated case where $|t_2 - t_1| \gg \tau_\mathrm{SD}$, which translates into bunching.

\begin{figure}[b]
  \centering
  \includegraphics[width=0.7\linewidth]{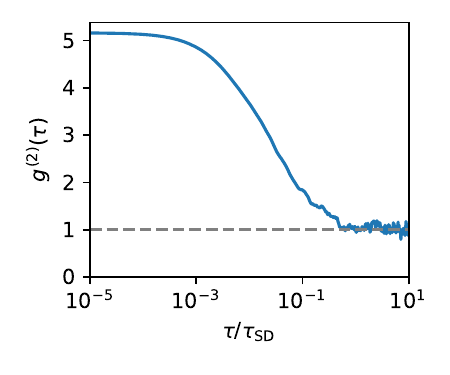}\\
  \caption{Example of $g^{(2)}(\tau)$ generated from a time trace shown Fig.~\ref{fig3}. The gray dashed line depicts the Poissonian limit $g^{(2)}(\tau) = 1$.}\label{fig4}
\end{figure}

Fig.~\ref{fig4} shows an example of $g^{(2)}$ function generated with a OU process. As expected, we observe $g^{(2)}(0) > g^{(2)}(\tau)$, indicating a bunching behavior resulting from the intensity fluctuations that originate from the spectral diffusion. The asymptotic limit $g^{(2)}(t \rightarrow \infty)$ is equal to~1 in the absence of additional random processes affecting the intensity stability, such as blinking. We define the degree of bunching $B(\tau) = g^{(2)}(\tau) - 1$ and the normalized degree of bunching $\tilde{B}(\tau) = B(\tau)/B(0)$, which decays from 1 to 0 as $\tau$ increases. We also define the bunching time $\tau_b$ such as $\tilde{B}(\tau_b) = 1/e$. In the following, we will establish the relation between $\tau_b$ and $\tau_\mathrm{SD}$ for the two classes of SD processes, after having reviewed some useful properties of the short-time bunching $B(0)$.

\subsubsection{1. Short-time bunching}

The short-time bunching can be deduced from the steady-state distribution of the emitter frequencies, independently of the underlying spectral diffusion model~\cite{Fournier23PRB}:

\begin{equation}
g^{(2)}(0) = \dfrac{\langle I^2 \rangle}{\langle I \rangle^2} = \dfrac{\int d \omega \mathcal{P}(\omega)C_{ss}(\omega)^2}{\left[\int d \omega \mathcal{P}(\omega)C_{ss}(\omega)\right]^2}
 \label{bunching}
\end{equation}
where $\mathcal{P}(\omega)$ is the PDF associated with the inhomogeneous distribution. In the case of an inhomogeneous linewidth largely exceeding the natural linewidth, this further simplifies to:

\begin{align}
g^{(2)}(0) &= \dfrac{1}{\mathcal{P}(\omega_{L})} \dfrac{\int d \omega C_{ss}(\omega)^2}{\left[\int d \omega C_{ss}(\omega)\right]^2} \nonumber \\
& =\dfrac{1}{\mathcal{P}(\omega_{L})} \dfrac{1}{\pi \Delta \omega_\mathrm{hom} }
 \label{bunching2}
\end{align}
where we have used Eq.~\ref{spectrum} to express the count rate $C_{ss}$ of a resonantly driven two-level system. When the PDF $\mathcal{P}(\omega_{L})$ is a Gaussian distribution, as is the case in the two models considered, we then obtain

\begin{equation}
g^{(2)}(0) = \dfrac{1}{2 \sqrt{\pi \ln 2}} 
\dfrac{\Delta \omega_\mathrm{inhom}}
{\Delta \omega_\mathrm{hom} } \exp\left[4 \ln 2 \left( \dfrac{\omega_L - \omega_0}{\Delta \omega_\mathrm{inhom}} \right)^2\right]
\label{bunching3}
\end{equation}
where $\omega_0$ denotes the center of the inhomogeneous distribution and $\Delta \omega_\mathrm{inhom}$ its full width at half maximum ($\Delta \omega_\mathrm{inhom} = 2 \sqrt{2 \ln2}\Sigma$). { At zero detuning, we then simply obtain $g^{(2)}(0) =  \frac{1}{2 \sqrt{\pi \ln 2}} 
\frac{\Delta \omega_\mathrm{inhom}}
{\Delta \omega_\mathrm{hom} }$.  This expression can be interpreted intuitively since, in two-photon correlations, the first detection informs that the spectrally diffusing emitter is close to resonance -- which at random times has a likelihood given by the ratio of homogeneous to inhomogeneous linewidth. This leads to an increased likelihood of detecting a second photon, which is maintained until spectral diffusion shifts the emitter spectral position out of resonance.} Remarkably, this result shows that it is possible to infer the homogeneous linewidth --and therefore the coherence time of the emitter-- from a simple measurement of the inhomogeneous (static) lineshape and the amount of bunching at a fixed laser detuning. { Moreover, this estimation does not require any assumptions about the nature of the microscopic mechanism yielding SD.} This result is all the more interesting as it is generally difficult to extract information about homogeneous properties in inhomogeneously broadened quantum systems, which can require complex experimental procedures such as spectral hole burning~\cite{Palinginis03} and dynamical decoupling~\cite{Abella66, Press10}.  { We note that, while Eq.~\ref{bunching3} is derived in the limit of Gaussian inhomogeneous spectra, Eq.~\ref{bunching} provides $\Delta \omega_\mathrm{hom}$ in the general case, where any experimental time-averaged lineshape can be accounted for numerically. We further discuss some of these considerations in the Supplemental Material (Sec.~S2).}

\begin{figure}[h]
  \centering
  \includegraphics[width=0.8\linewidth]{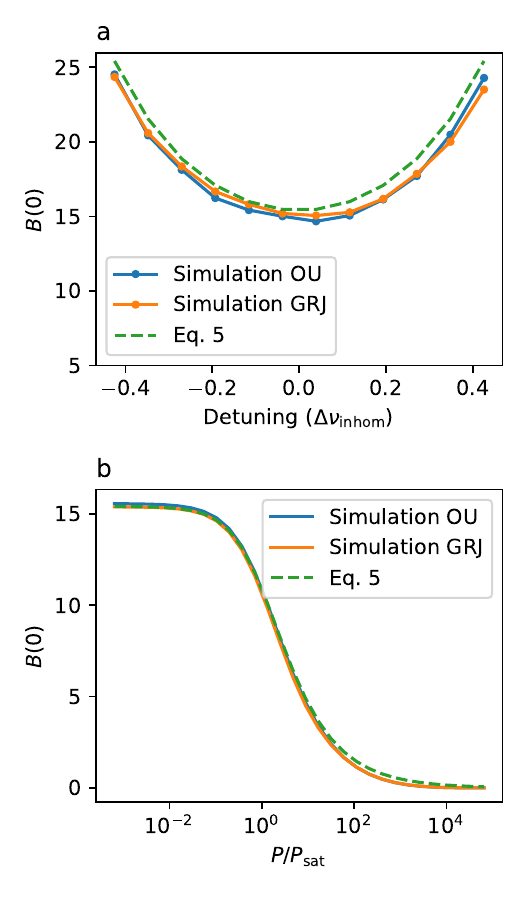}\\
  \caption{(a) Zero-time bunching $B(0)$ as a function of the laser detuning. (b) $B(0)$ as a function of the laser power. In both simulations, $\Delta \omega_\mathrm{hom}/\Delta \omega_\mathrm{inhom} = 2.2 \cdot 10^{-2}$.  }\label{fig5}
\end{figure}

To illustrate the dependence of the short-time bunching on the homogeneous linewidth, on Fig.~\ref{fig5}a we plot $B(0)$ as a function of the laser power. The decrease of the amount of bunching as power increases, already observed experimentally~\cite{Fournier23PRB}, is a consequence of power broadening yielding a larger homogeneous response, as discussed in the previous section. The dependence of $B(0)$ on the laser detuning is plotted on Fig.~\ref{fig5}a. In this situation, bunching increases with the detuning due to the decreasing likelihood for the emitter to be resonant with the laser when the detuning increases. { More examples of calculations are given in the Supplemental Material (Sec.~S2).} In both cases, we verify numerically that, as expected, the observed behavior is independent of the SD mechanism and correlation time. This is however not the case of the bunching decay, as we expose in the following section.

\subsubsection{2. Bunching decay}

\begin{figure}[h]
  \centering
  \includegraphics[width=0.9\linewidth]{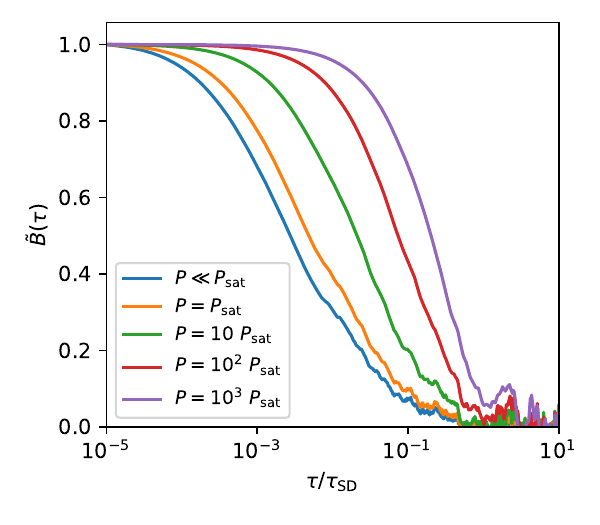}\\
  \caption{Normalized bunching $\tilde{B}(\tau)$ as a function of $\tau$ in units of $\tau_\mathrm{SD}$ for various laser powers.  }\label{fig6}
\end{figure}

Since the amount of bunching $B(0)$ is independent of the model and only depends on the laser detuning and power, we now focus on the normalized bunching $\tilde{B}(\tau)$, and we fix the laser frequency to the center of the inhomogeneous distribution ($\omega_L = \omega_0$). We first investigate the normalized intensity correlation function in the case of a continuous spectral diffusion governed by a OU process. Fig.~\ref{fig6} shows a numerical calculation of $\tilde{B}(\tau)$ for various laser powers above saturation. Two important characteristics can be observed: firstly, the decay is not exponential. Secondly, the characteristic time $\tau_b$ increases with the power. This latter observation demonstrates that the bunching timescale $\tau_b$ is in general not equal to the correlation time $\tau_\mathrm{SD}$. Since the considered model is continuous, $\tau_b$ represents the characteristic time during which the emitter stays near resonance with the laser before spectral diffusion detunes it significantly. In presence of power broadening, it is therefore expected that the increased linewidth prolongs this duration when the laser power increases.

To gain more quantitative insight, we calculate the correlation function analytically, based on the following expression: 
\begin{equation}
g^{(2)}(\tau) = \dfrac{1}{N} \iint d\omega d\omega' \mathcal{P}(\omega) C_{ss}(\omega)  P(\omega', t + \tau | \omega, t) C_{ss}(\omega')
 \label{g2_OU}
\end{equation}
where $N = [\int \mathcal{P}(\omega) d\omega C_{ss}(\omega)]^2$. We have introduced the spectral diffusion kernel~\cite{Risken89}

\begin{equation}
P(\omega', t' | \omega, t) = \dfrac{\exp\left(- \dfrac{\left(\omega' - \omega e^{-(t'-t)/\tau_\mathrm{SD}}\right)^2}{2 \Sigma^2 (1 - e^{-2(t'-t)/\tau_\mathrm{SD}})} \right)}{\sqrt{2 \pi} \Sigma \sqrt{1 - e^{-2(t'-t)/\tau_\mathrm{SD}}}}
\end{equation}
which expresses the conditional probability density function for the emitter frequency to be found at position $\omega'$ at time $t'$ given position $\omega$ at time $t$. In the limit where $\Delta \omega_\mathrm{inhom} \gg \Delta \omega_\mathrm{hom}$, the intensity correlation then reads

\begin{equation}
g^{(2)}(\tau) = \dfrac{1}{2\sqrt{\pi \ln 2}}\dfrac{\Delta \omega_\mathrm{inhom}}{\Delta \omega_\mathrm{hom}}f \left( \dfrac{ 2 \Delta \omega_\mathrm{hom} \sqrt{\ln 2}}{\Delta \omega_\mathrm{inhom} \sqrt{1 -  e^{-2\tau/\tau_\mathrm{SD}}}}\right)
\label{g2_ou_analytical}
\end{equation}

with 

\begin{equation}
f(x) = \sqrt{\pi} x e^{x^2} \mathrm{erfc}(x)
\end{equation}

\begin{figure}[b]
  \centering
  \includegraphics[width=0.9\linewidth]{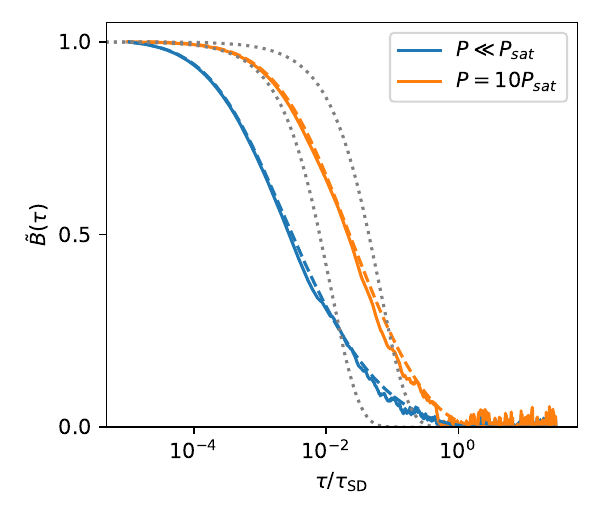}\\
  \caption{Plain curves (dashed curves): numerical simulation (analytical calculation using Eq.~\ref{g2_ou_analytical}) of $\tilde{B}(\tau)$ as a function of $\tau$ in units of $\tau_\mathrm{SD}$ for two laser powers. Dotted gray lines: exponential fits of the simulations.}\label{fig7}
\end{figure}

Fig.~\ref{fig7} displays the analytical result of Eq.~\ref{g2_ou_analytical} together with the numerical simulation for two different powers (and therefore two different values of $\Delta \omega_\mathrm{hom}/\Delta \omega_\mathrm{inhom}$). An exponential fit is also plotted to indicate the clear deviation from the exponential behavior. Based on Eq.~\ref{g2_ou_analytical}, we can infer an approximate expression for $\tau_b$ in the limit $\Delta \omega_\mathrm{inhom} \gg \Delta \omega_\mathrm{hom}$, which provides
\begin{equation}
\tau_b = K \left(\dfrac{\Delta \omega_\mathrm{inhom}}{\Delta \omega_\mathrm{hom}} \right) ^2 \tau_\mathrm{SD}
\label{tau_b_OU}
\end{equation}
with $K \approx 18$. This expression for $\tau_b$ can be understood intuitively. At short times, the OU process is equivalent to the Wiener process of Brownian motion. Therefore, an emitter close to resonance at $t = 0$ probes a spectral range growing as $\sigma(t) \approx \Sigma \sqrt{t/\tau_\mathrm{SD}}$. It then escapes resonance when the standard deviation of spectral diffusion becomes comparable to its homogeneous linewidth, \textit{i.e.} for $\sigma(t = \tau_b) \sim \Delta \omega_\mathrm{hom}$, yielding $\tau_b \propto \tau_\mathrm{SD}\left(\Delta \omega_\mathrm{hom} / \Delta \omega_\mathrm{inhom} \right)^2$. 

\begin{figure}[t]
  \centering
  \includegraphics[width=0.9\linewidth]{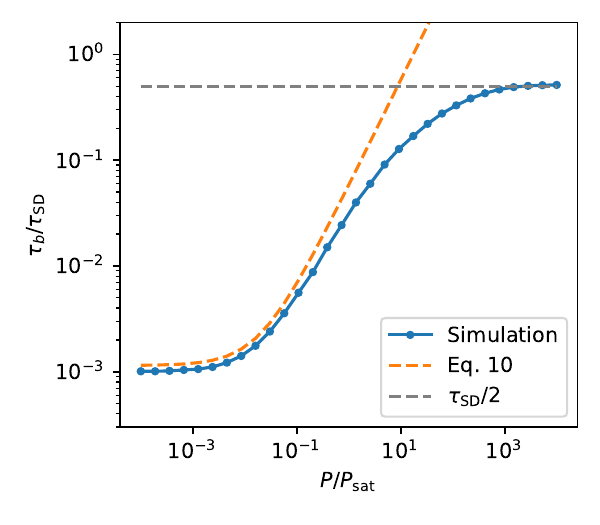}\\
  \caption{Bunching time $\tau_b$ as a function of the laser power for $\Delta \omega_\mathrm{hom} / \Delta \omega_\mathrm{inhom} \approx 130$. Blue line: numerical simulation; orange dashed line: calculation from Eq.~\ref{tau_b_OU}; gray dashed line: high power asymptote $\tau_b = \tau_\mathrm{SD}/2$.}\label{fig8}
\end{figure}

Fig.~\ref{fig8} shows the bunching time $\tau_b$ as calculated numerically as a function of the laser power. At low powers, it is constant, given by Eq.~\ref{tau_b_OU}. Above saturation, $\tau_b$ increases due to power broadening prolonging the duration of bright periods, and then saturates when the homogeneous linewidth exceeds the width of the inhomogeneous distribution, with in this case $\tau_b = \tau_\mathrm{SD}/2$. The factor~1/2 originates from the fact that, in the limit $\Delta \omega_\mathrm{hom} \gg \Delta \omega_\mathrm{inhom}$, the system is first-order insensitive to SD~\cite{Kuhlmann13}. Therefore, at the second order, fluctuations of $\omega(t)$ at a frequency $f$ yield fluctuations of $C(t)$ at a frequency $2f$ and thus the associated decay time is halved.

\begin{figure}
  \centering
  \includegraphics[width=0.9\linewidth]{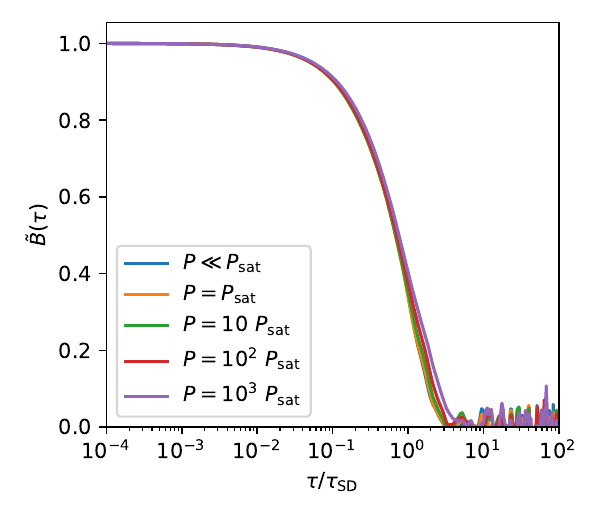}\\
  \caption{Normalized bunching $\tilde{B}(\tau)$ as a function of $\tau$ in units of $\tau_\mathrm{SD}$ for various laser powers.  }\label{fig9}
\end{figure}

We now consider the GRJ model. Fig.~\ref{fig9} shows the numerically calculated $\tilde{B}(\tau)$ at various laser powers. Contrarily to the OU case, here the bunching decay time and shape do not vary with the power. This can be seen intuitively: if the emitter is on resonance with the laser at $t = 0$, it stays so in average during $\tau_\mathrm{SD}$, independently of the laser power, before a spectral jump detunes the emitter transition energy. The laser power only acts on the bare probability to be close to resonance at a given time -- and therefore on $B(0)$, as discussed in the previous section. As a consequence, $\tilde{B}(\tau)$ directly inherits its time dependence from the SD correlation function $\langle \omega(t+\tau)\omega(t) \rangle$ and exhibits an exponential decay of characteristic time  $\tau_b = \tau_\mathrm{SD}$, as verified in our simulations.

These observations establish that the long-time $g^{(2)}(\tau)$ carries clear, experimentally accessible signatures of the underlying spectral diffusion stochastic process: in the case of the OU model, $g^{(2)}(\tau)$ decays non-exponentially according to Eq.~\ref{g2_ou_analytical}, and the decay time increases with the laser power above saturation. In the case of a GRJ mechanism, $g^{(2)}(\tau)$ decays exponentially and the bunching time is independent of the power. Only in the latter case do we have $\tau_b = \tau_\mathrm{SD}$.

{ We observe that a power-independent decay of $g^{(2)}(\tau)$ is a general signature of discrete jumps of the emitter to uncorrelated spectral positions within the inhomogeneous distribution. The decay of $g^{(2)}(\tau)$ is then directly inherited from the time statistics of the jumps, and is exponential in the case of Poissonian jumps. On the opposite, a power-dependent decay time indicates that intermediate spectral positions of the emitter are probed increasingly as power broadens the emitter, and therefore constitutes a general signature of continuous spectral diffusion mechanisms.

In the Supplemental Material (Sec.~S4 and~S5), we provide examples of emitters subject to more complex mechanisms, such as a combination of two SD components, as well as a contribution from blinking -- showing thereby that our approach can be in principle generalized to a wide range of experimental situations.}

\subsection{B. Intensity fluctuations}

{ In this section, we investigate additional consequences of SD on the photon statistics of resonance fluorescence. We show that the long-time photon statistics are affected by SD, in a way that is inherited from the microscopic model.}

\subsubsection{1. Standard deviation of the intensity distribution}

At a fixed laser detuning, the average intensity --or count rate-- that the emitter scatters is simply given by $\langle C(\omega_L) \rangle = \int d\omega \mathcal{P}(\omega) C(\omega)$. The count rate is typically measured by integrating the photon number over a macroscopic time $T$. This measured intensity fluctuates because of shot noise, with a relative intensity noise given by $\Delta I_T / \langle I \rangle_T = 1 / \sqrt{\langle N \rangle_{T}}$, where $\langle N \rangle_{T}$ is the average number of photons detected during ${T}$. These Poissonian fluctuations can be mitigated by optimizing the photon collection efficiency. Additional sources of noise can cause the variance $V_T$ of the photon number during $T$ to exceed the Poisson limit $V_T = \langle N \rangle_{T}$. The Mandel parameter $Q = V_T/\langle N \rangle_{T} - 1$ relates the long-time intensity fluctuations to the second-order intensity correlation via the following relation~\cite{Mandel79,Short83,Treussart02,Canneson14}
\begin{equation}
Q =\dfrac{\langle N \rangle_{T}}{T} \int_{-T}^{T} d\tau
\left(
1 - \dfrac{|\tau|}{T}
\right)
\left( 
g^{(2)}(\tau) - 1
\right)
\label{Mandel}
\end{equation}

If $g^{(2)}(\tau) =1$, we have $Q = 0$ and the intensity noise is given by the shot noise. As can be seen from Eq.~\ref{Mandel}, the presence of finite bunching increases the intensity noise above the Poisson limit. Since spectral diffusion is associated with photon bunching as discussed in the previous sections, it is also expected to yield additional intensity fluctuations at macroscopic times ($T \gg \tau_\mathrm{SD}$).

To isolate the contribution of spectral diffusion to the intensity noise, we now consider the high intensity (or high photon collection) limit where the shot noise can be neglected ($\langle N \rangle_T \gg 1$). Since the latter is statistically independent from the SD noise, it can be reintegrated at a later step { (see Supplemental Material Sec.~S3)}. In this limit, the relative intensity fluctuations take the form
\begin{equation}
\dfrac{\Delta I_T}{\langle I \rangle_T} = \left[\dfrac{1}{T} \int B(\tau)d\tau\right]^{\frac{1}{2}}
\end{equation}
which, when the laser is fixed at the center of the inhomogeneous distribution, writes
\begin{equation}
\dfrac{\Delta I_T}{\langle I \rangle_T} = \left[\dfrac{1}{\sqrt{\pi \ln 2 } T}\dfrac{\Delta \omega_\mathrm{inhom}}{\Delta \omega_\mathrm{hom}} \int \tilde{B}(\tau)d\tau\right]^{\frac{1}{2}}
\label{Ifluct2}
\end{equation}

In the case of GRJ model, Eq.~\ref{Ifluct2} further simplifies to
\begin{equation}
\dfrac{\Delta I_T}{\langle I \rangle_T} = \sqrt{\dfrac{1}{\sqrt{\pi \ln 2 }} \dfrac{\Delta \omega_\mathrm{inhom}}{\Delta \omega_\mathrm{hom}} \dfrac{\tau_\mathrm{SD}}{T}}
\label{IfluctGRJ}
\end{equation}

\begin{figure}[h]
  \centering
  \includegraphics[width=0.9\linewidth]{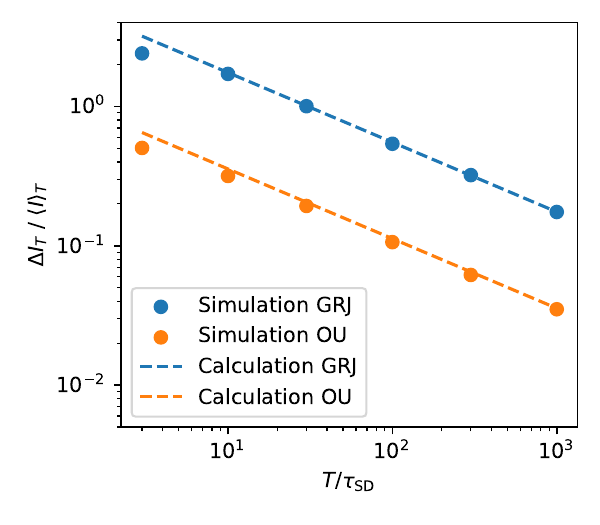}\\
  \caption{Relative intensity noise as a function of the integration time $T$. Blue (orange) dots: simulations for the GRJ (OU) model. Blue (orange) dashed lines: calculation for the GRJ (OU) model using Eq.~\ref{IfluctGRJ} (Eq.~\ref{Ifluct2}). }\label{fig10}
\end{figure}

In the case of SD based on OU process, the integral in Eq.~\ref{Ifluct2} has, {to the best of our knowledge, }no analytical solution. We therefore perform a numerical evaluation of this expression. Fig.~\ref{fig10} shows the result of these calculations together with a numerical simulation for varying integration time, at low power. The simulations and the analytical calculations are in good agreement, in particular for $T\gg\tau_\mathrm{SD}$. The intensity fluctuations for the OU process are smaller than in the GRJ case, due to the fact that $B(\tau)$ decays much faster in the former case. On the other hand, both have identical dependence on the integration time $T$. Therefore the intensity noise at a given power is not sufficient to identify the SD process without a prior knowledge of $\tau_\mathrm{SD}$.

\begin{figure}[h]
  \centering
  \includegraphics[width=0.8\linewidth]{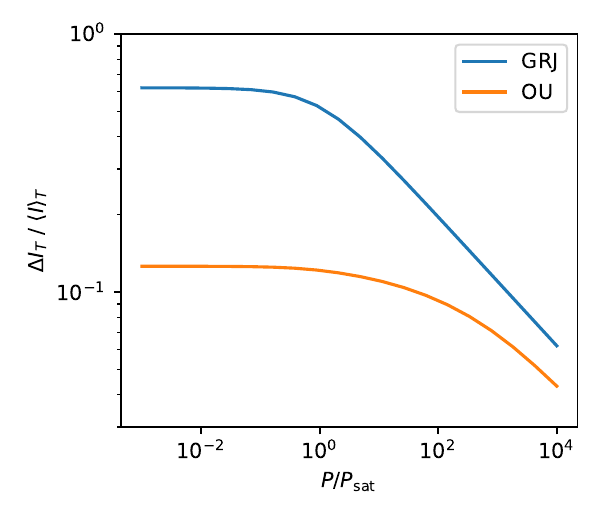}\\
  \caption{Relative intensity noise as a function of the excitation laser power.}\label{fig11}
\end{figure}

Fig.~\ref{fig11} shows the power dependence of the relative intensity noise, as calculated from Eq.~\ref{Ifluct2}. In both cases, the fluctuations decrease with increasing power. However, in the case of the GRJ model, the reduction occurs around the saturation power, while in the OU model, the reduction occurs at a power such that $\Delta\omega_\mathrm{hom} \sim \Delta\omega_\mathrm{inhom}$, which is much higher in many practical situations where $\Delta\omega_\mathrm{inhom} \gg 1/T_2$. The example of Fig.~\ref{fig11} is taken with $\Delta\omega_\mathrm{inhom} / \Delta\omega_\mathrm{hom} \approx 50$, leading to a two-fold decrease at $P \approx 10^3 P_\mathrm{sat}$ in the OU case. Therefore, knowing the saturation power, the power dependence of the intensity fluctuations constitutes a macroscopic-time manifestation of the underlying microscopic SD mechanism.

\subsubsection{2. Skewness of the intensity distribution}

Another informative aspect carried in the long-time intensity fluctuations is the skewness of the intensity distribution. The skewness quantifies the asymmetry of a distribution and is defined as
\begin{equation}
\gamma_1 = \left\langle \left( \dfrac{I - \langle I \rangle_T}{\Delta I_T} \right)^3 \right\rangle
\end{equation}

\begin{figure}[h]
  \centering
  \includegraphics[width=0.8\linewidth]{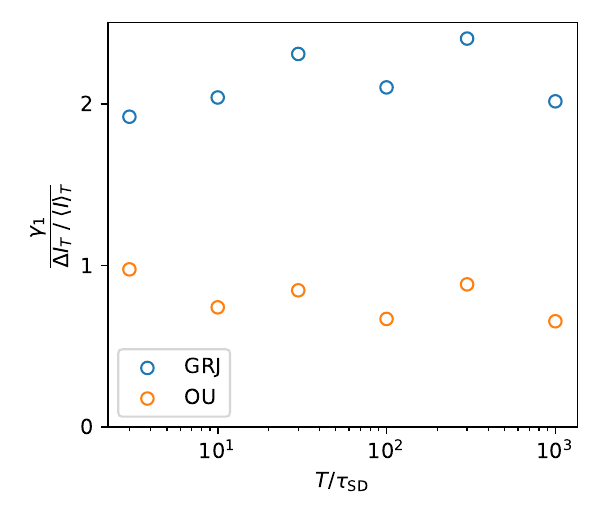}\\
  \caption{Relative skewness as a function of the integration time. }\label{fig12}
\end{figure}

For a Poisson distribution, $\gamma_1 = \Delta I_T/\langle I \rangle_T$. Fig.~\ref{fig12} plots the skewness of the time trace used in the previous section, normalized by the intensity noise. In the OU model, we obtain a close to Poissonian skewness $\gamma_1 \approx \Delta I/I$ for all bin sizes. On the contrary, in the GRJ model, we observe super-Poissonnian skewness $\gamma_1 \approx 2 \Delta I/I$, demonstrating an additional impact of the SD mechanism on the statistics of the integrated intensity.

\begin{figure}[h]
  \centering
  \includegraphics[width=0.8\linewidth]{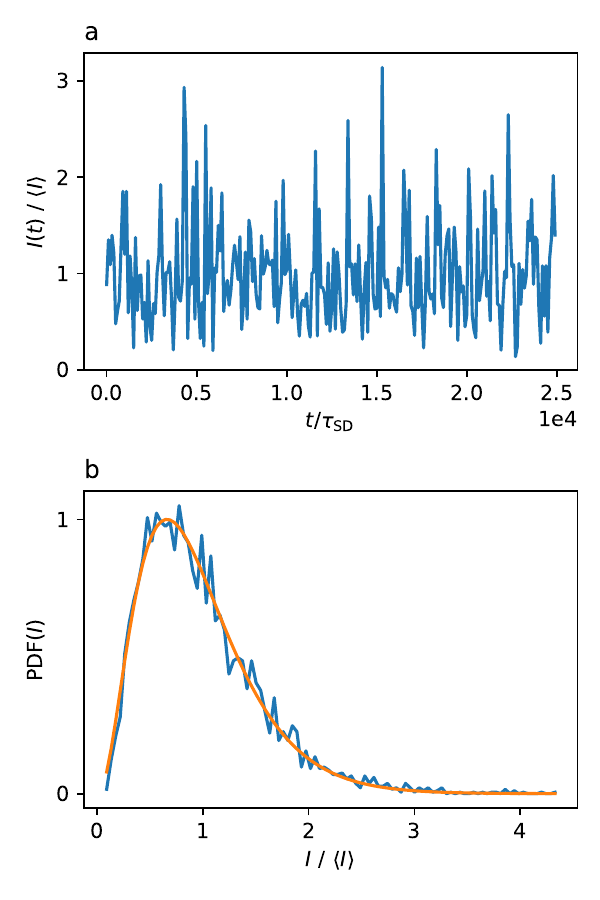}\\
  \caption{(a) Intensity time trace with integration time $T = 100 \tau_\mathrm{SD}$. (b) Blue curve: histogram of the intensities. Orange curve: fit using a gamma distribution. }\label{fig13}
\end{figure}

\begin{figure}[h]
  \centering
  \includegraphics[width=0.8\linewidth]{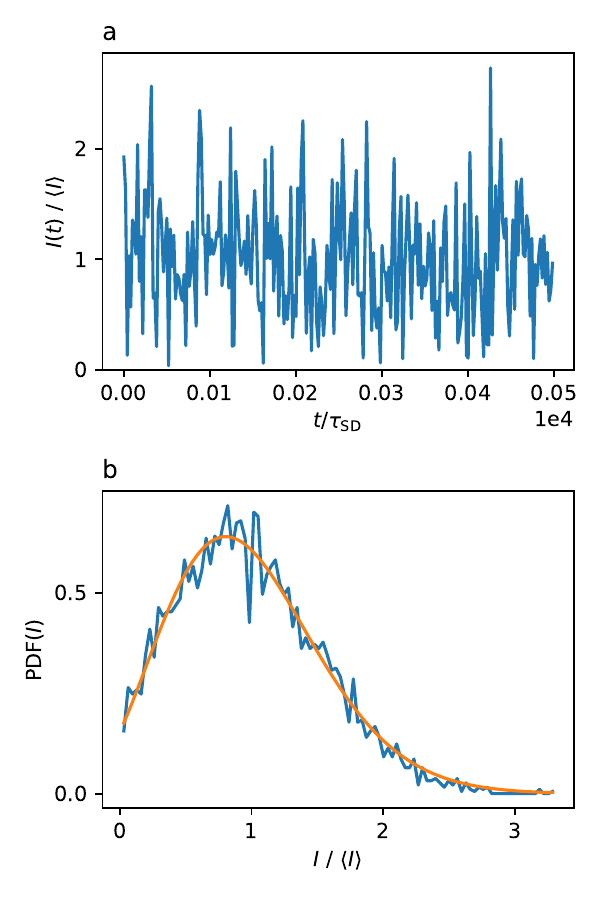}\\
  \caption{(a) Intensity time trace generated using the OU model. (b) Blue curve: histogram of the intensities. Orange curve: fit using a Poisson distribution. }\label{fig14}
\end{figure}

The super-Poissonian skewness of the intensity distribution in the case of the GRJ model can be well understood considering that the duration of the stable periods --as defined as the time between two jumps-- is exponentially distributed, an elementary property of Poisson processes. During the time $T_\mathrm{bin}$, there are in average $T_\mathrm{bin}/\tau_\mathrm{SD}$ stable periods. The fraction of them which are bright (\textit{i.e.} $\omega \approx \omega_L$) is $\sim \Delta \omega_\mathrm{hom}/2 \sqrt{\pi \ln 2} \Delta \omega_\mathrm{inhom}$, where we have supposed a Gaussian PDF and zero detuning ($\omega_L = \omega_0$). The average number of bright stable periods during $T_\mathrm{bin}$ is $N_b = (T_\mathrm{bin}/\tau_\mathrm{SD})\Delta \omega_\mathrm{hom}/2\sqrt{\pi \ln 2}\Delta \omega_\mathrm{inhom}$. By making the approximation that the number of bright periods is independent of their duration --which is only valid asymptotically--, the intensity can then be written as a sum of exponentially distributed random variables. Therefore, the intensity is itself a random variable whose distribution is a gamma distribution, of parameters $N_b$ and $\tau_c$. This distribution has a skewness of $\gamma_1 = 2 \Delta I/I$. Fig.~\ref{fig13}a shows an intensity trace, where the occurrence of short periods of brightness sizably higher than average can be observed as a manifestation of the asymmetry of the intensity distribution. Fig.~\ref{fig13}b shows the histogram of the integrated intensities over a long period, together with a fit using the gamma distribution, showing the excellent agreement with the simulation. {Such histogram can provide an estimation of the spectral diffusion time, based on prior knowledge of $\Delta \omega_\mathrm{hom} / \Delta \omega_\mathrm{inhom}$. By using Eq.~\ref{IfluctGRJ} together with the relation $\gamma_1 = 2 \Delta I/I$ valid for any gamma distribution, we obtain $\tau_\mathrm{SD} = \sqrt{\pi \ln 2} \frac{\Delta\omega_\mathrm{inhom}}{\Delta\omega_\mathrm{hom}} T (\gamma_1/2)^2$, which we have verified provides back the input value used in the simulation, within 10~\%. For comparison, we also simulate a time trace (shown Fig.~\ref{fig14}a) and the corresponding intensity histogram (Fig.~\ref{fig14}b) in the case of the OU model. The latter can be best fitted with a Poisson distribution, consistently with the Poissonian skewness discussed above. Note the non-zero intercept in this case as opposed to the GRJ case.} These observations suggest that measurement of super-Poissonian skewness in intensity traces provides evidence for a jump-based spectral diffusion process, as well as an estimation of the spectral diffusion time.

\section{IV. Discussion and conclusion}

We have provided a general framework to describe and analyze the photon statistics of spectrally diffusive emitters resonantly driven by a cw laser. Our simulations, supported by analytical calculations, have evidenced several experimentally accessible signatures of the discrete or continuous nature of the stochastic process governing the variations of the emitter transition energy in time. Two of these characteristic signatures are associated with the intensity correlation function $g^{(2)}$. {As a general rule, a decrease in the amount of bunching that is not accompanied by an increase in the bunching decay time is characteristic of discrete jumps to uncorrelated positions.}In this regard, a recent experimental study has established the presence of power-dependent bunching in the $g^{(2)}$ of a B center in hBN~\cite{Fournier23PRB}. In light of the results exposed in the present work, we can attribute the SD mechanism of B centers to a discrete jump process, owing to the fact that the bunching timescale is almost independent of the power above saturation (up to $P \approx 100 P_\mathrm{sat}$). This observation is consistent with what has been demonstrated with other color centers in hBN based on photon correlation Fourier spectroscopy~\cite{Spokoyny20}. In the latter reference, the GRJ process is evidenced by a direct measurement of the spectral correlation function, which would not be possible in the case of B centers due to their much narrower homogeneous and inhomogeneous widths. As a property of the GRJ model that we have established in section~III, the bunching time is equal to the spectral diffusion time, which in the case of B centers has been measured to be $\sim 30$~$\mu$s~\cite{Fournier23PRB}. The present work provides guidelines for a more in-depth investigation of spectral diffusion in these systems, as well as in a large variety of other single-photon sources. It can be readily extended to other potential microscopic models, as well as to other experimental processes. As an example, an alternative method that does not require resonant excitation is to filter the emitter fluorescence with a bandwidth narrower than the inhomogeneous linewidth. In the Supplemental Material (Sec.~S1), we investigate the statistical properties of light filtered from non-resonantly excited emitters. We show that it shares some similarities with resonance fluorescence, where the spectral selectivity is implemented by the resonant laser. It therefore carries analogous signatures. {For instance, observation of a bunching time independent of the filter width implies discrete jumps to the whole inhomogeneous distribution. In addition,} the bunching time associated with continuous diffusion is also shorter than the SD time, which possibly modifies the conclusions of prior work using this technique~\cite{Abbarchi12}. {We also present examples of more complex situations, where the emitter subject to blinking (Sec.~S4), or where the emitter subject to more than one SD mechanism (Sec.~S5) illustrating thereby the versatility of our approach. } 

Given the ubiquity of spectral diffusion processes in condensed matter as well as the importance of resonance fluorescence in the study and control of solid-state quantum emitters, we expect our work to play a helpful role in the general understanding and technological development of quantum dots and color centers, with applications to quantum computing and quantum networks.

\section{Acknowledgments}
The authors thank Clarisse Fournier for fruitful discussions. This work is supported by the French Agence Nationale de la Recherche (ANR) under reference ANR-21-CE47-0004-01 (E$-$SCAPE project).

pagebreak
~
\newpage

\onecolumngrid
\begin{center}
  \textbf{\large Supplemental Material\\~\\Photon statistics of resonantly driven spectrally diffusive quantum emitters}\\[.2cm]
  Aymeric Delteil, St\'ephanie Buil, Jean-Pierre Hermier\\[.1cm]
  {\itshape \small Universit\'e Paris-Saclay, UVSQ, CNRS,  GEMaC, 78000, Versailles, France. \\
{\color{white}--------------------} aymeric.delteil@usvq.fr{\color{white}--------------------} \\}

\end{center}

\setcounter{equation}{0}
\setcounter{figure}{0}
\setcounter{table}{0}
\setcounter{page}{1}
\renewcommand{\theequation}{S\arabic{equation}}
\renewcommand{\thefigure}{S\arabic{figure}}

\section{S1. Filtered spontaneous emission}

Under non-resonant excitation, the emitted light does not exhibit bunching in the absence of additional processes causing intensity fluctuations, such as blinking. However, if a narrowband filter is placed on the detection path, spectral diffusion brings the emitter in and out of resonance with the filter, yielding intensity fluctuations after the  filter. This technique shares some similarities with resonant excitation, on account of a spectrally selective process converting spectral variations into intensity fluctuations, which can be then analyzed using photon counting techniques. It has been used for instance to study SD in self-assembled quantum dots~\cite{Abbarchi12}. For these reasons, in this section we investigate the statistical properties of filtered spontaneous emission of a spectrally diffusive emitter subject to either previously described SD mechanism.

\begin{figure}[h]
  \centering
  \includegraphics[width=0.45\linewidth]{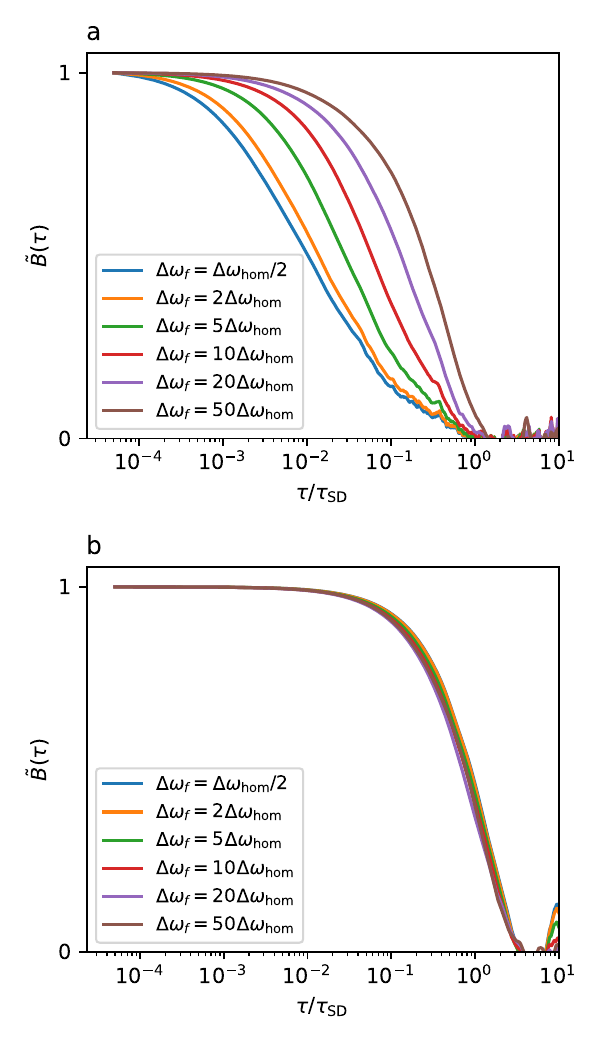}\\
  \caption{Bunching function $\tilde{B}(\tau)$ for increasing filter bandwidth for (a) the OU model and (b) the GRJ model.}\label{figS1}
\end{figure}

\begin{figure}[h]
  \centering
  \includegraphics[width=0.45\linewidth]{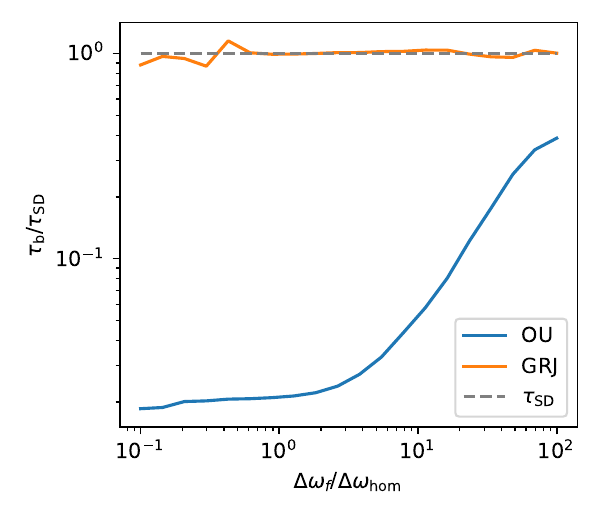}\\
  \caption{Bunching time $\tau_b$ as a function of the filter width.}\label{figS2}
\end{figure}

We consider a rectangular filter of bandwidth $\Delta \omega_f$. We generate time traces from a spectral trajectory $\omega(t)$ by recording the overlap between the Lorentzian spectrum centered at $\omega(t)$ and the rectangular filter of which we fix the center to $\omega_0$, the center of the inhomogeneous distribution. We then calculate the intensity correlation function based on eq.~2 of the main text. Figure~\ref{figS1} shows the resulting bunching function $\tilde{B}(\tau)$ for both SD models for different values of $\Delta \omega_f$. In the case of the OU continuous diffusion, the decay is non-exponential and varies with the filter width. On the contrary, in the case of the GRJ model, the decay is exponential with a fixed decay time. This is reminiscent of the resonant excitation case. The difference is that here, the selectivity of the process yielding detuning-dependent intensity is tuned using the filter bandwidth instead of the laser power. Therefore, many of the results of the main text section~III can be transposed to this configuration. As an example, figure~\ref{figS2} shows the dependence of the decay time $\tau_b$ on the filter width. In the case of the OU model, the decay time is much shorter than $\tau_\mathrm{SD}$ and increases with the filter width -- in particular when the latter exceeds the homogeneous linewidth. On the other hand, in the GRJ case, the bunching decay time is fixed to $\tau_\mathrm{SD}$, as in the case of resonant excitation.

The reason for such behavior is similar: observation of a bunching time that independent of the filter width implies jumps from within the filter bandwidth to new uncorrelated positions within the whole inhomogeneous distribution. Otherwise, more and more intermediate positions would be captured as the filter bandwidth bandwidth is increased, yielding bandwidth-dependent bunching time. 

Experimentally, a measurement of the bunching time with at least two different filter widths would allow to infer the continuous nature of the spectral diffusion. In reference~\cite{Abbarchi12}, since only one narrow filter was used to infer the bunching time, the study cannot be conclusive in this respect -- and therefore the observed bunching time cannot be matched with the spectral diffusion time.

\clearpage

\section{S2. Influence of the ratio $\Delta  \omega_\mathrm{hom}/\Delta  \omega_\mathrm{inhom}$ on zero-time bunching}

In this section, we provide more insight into the zero-time bunching investigated in section III. A 1 of the main text.

\subsection{A. Zero-time bunching at zero detuning as a function of the inhomogeneous broadening}

\begin{figure}[h]
  \centering
  \includegraphics[width=0.5\linewidth]{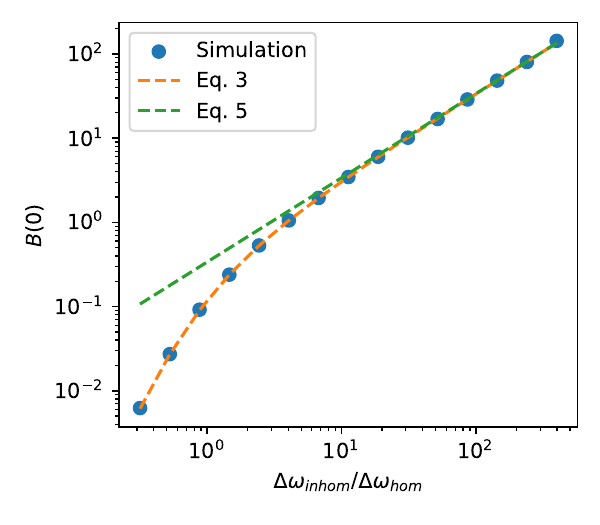}\\
  \caption{Zero-time bunching $B(0)$ as a function of $\mathrm{inhom} / \omega_\mathrm{hom}$. Blue dots: simulation. Dashed orange line: calculation based on Eq.~3. Dashed green line: calculation based on Eq.~5.}
  \label{zerotimebunching}
\end{figure}

Fig.~\ref{zerotimebunching} shows the zero-time bunching as a function of the ratio between the SD-induced broadening $\Delta \omega_\mathrm{inhom}$ and the homogeneous linewidth $\Delta \omega_\mathrm{hom}$, as calculated by the approximate formula given by Eq.~5, together with a simulation and the exact calculation given by Eq.~3, which is valid for $\Delta \omega_\mathrm{inhom} \gg\Delta  \omega_\mathrm{hom}$. The laser is fixed at the center of the inhomogeneous distribution. We can observe that, for large homogeneous distributions, both formulas agree with the simulations. However, in case of weak spectral diffusion ($\Delta \omega_\mathrm{inhom} \lesssim \Delta  \omega_\mathrm{hom}$), Eq.~5 overestimates the zero-time bunching (or equivalently, given a measurement of $B(0)$, it underestimates $\omega_\mathrm{hom}$). Eq.~3, in turn, succeeds in providing the correct linewidth ratio from the zero-time bunching. We emphasize that measuring bunching values of less than 1~\% is experimentally challenging. As we discuss in the next section, in the case of weakly broadened emitters, the optimal laser position to observe bunching is not at resonance but detuned by half the linewidth.

\subsection{B. Zero-time bunching as a function of the detuning for various inhomogeneous broadenings}

\begin{figure}[h]
  \centering
  \includegraphics[width=1.0\linewidth]{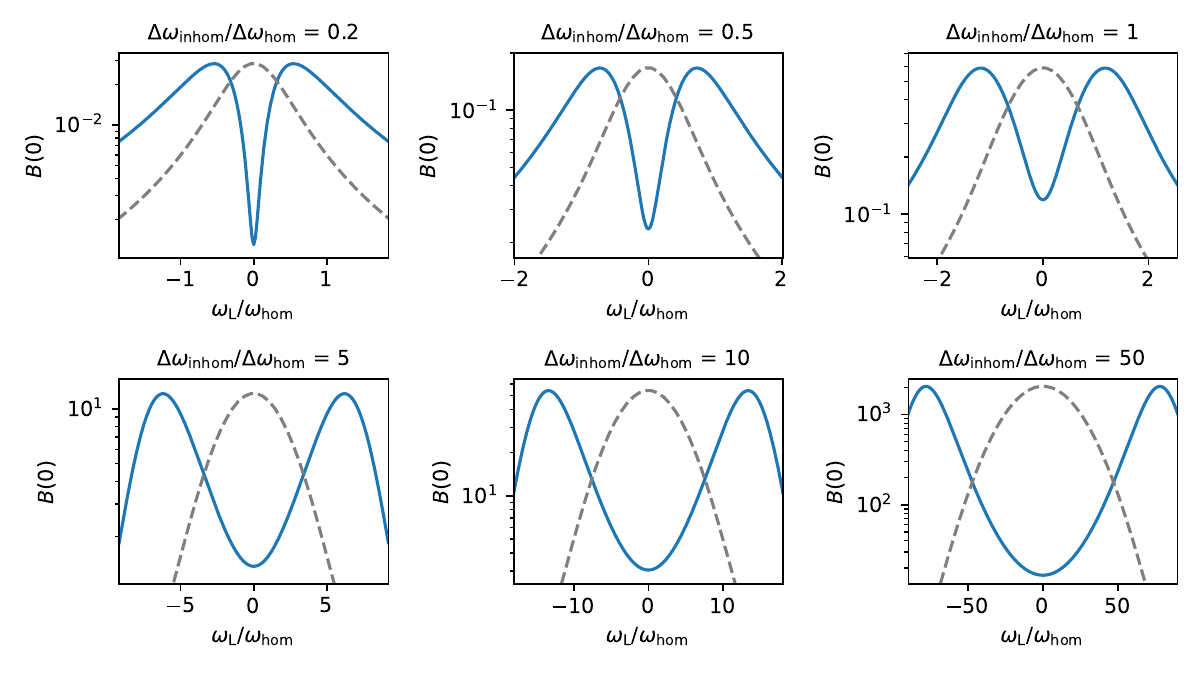}\\
  \caption{Zero-time bunching as a function of the laser detuning, for various values of the ratio $\Delta\omega_\mathrm{inhom}/\Delta\omega_\mathrm{hom}$ (blue curves). The fluorescence intensity is indicated in dashed gray lines, and evolves from Lorentzian to Gaussian.}\label{ztb_det}
\end{figure}

Fig.~\ref{ztb_det} shows the detuning dependence of the zero-time bunching $B(0)$, as calculated from Eq.~3, for various SD amplitudes ranging from weakly broadened emitters (such as in ref.~\cite{Kuhlmann13}) to strongly broadened emitters (such as in refs.~\cite{Fournier23PRB, Wolters13}). The detuning dependence shifts from a squared dispersive shape to an inverse Gaussian shape corresponding to the limit of Eq.~5. When the SD is weak, the fluorescence intensity is first-order insensitive to spectral fluctuations for exactly resonant laser drive (i.e. at zero detuning). In turn, the sensitivity is maximal at half-linewidth detuning -- an observation already made in prior work on weakly broadened QDs~\cite{Kuhlmann13}. On the other hand, in the case of strongly broadened emitters, the sensitivity sizably increases in the regions where the count rate vanishes. In this case, it is therefore more practical to perform experiments at resonance.

\clearpage

\section{S3. Shot noise contribution}

Any real-world experiment on single-photon emitters is affected by shot noise. In this section, we draw some general considerations on the level of signal needed to observe the signatures of spectral diffusion described in the main text. On the one hand, it is well known that shot noise is uncorrelated, such that it do not affect the shape of the $g^{(2)}$ function. Therefore, all SD signatures described section III A of the main text are unaffected by shot noise. However, this is not true for the long-time statistics we describe in section III B. In order to infer the relative role of shot noise as compared with SD-induced intensity fluctuations, we consider the GRJ model, which provides exact expressions for the photon statistics.
Based on Eq.~14, the intensity noise is given by 
\[
\left(\dfrac{\Delta I}{I}\right)_\mathrm{SD} = \sqrt{\dfrac{1}{\sqrt{\pi \ln 2}}  \dfrac{\Delta\omega_\mathrm{inhom}}{\Delta\omega_\mathrm{hom}}\dfrac{\tau_\mathrm{SD}}{T}  }
\]

where $T$ is the integration time (i.e. bin width).
The shot noise depends on the count rate as
\[
\left(\dfrac{\Delta I}{I}\right)_\mathrm{SN} = \dfrac{1}{\sqrt{RT}}
\]

where $R$ is the photon count rate. Therefore, having photon statistics dominated by SD-induced noise requires 
\[
R > R_\mathrm{eq} = \sqrt{\pi \ln 2}\dfrac{\Delta\omega_\mathrm{hom}}{\Delta\omega_\mathrm{inhom}} \dfrac{1}{\tau_\mathrm{SD}} 
\]

In the following table, we include a few numerical applications of various single-photon emitters from the literature.
\newline
\begin{center}
\begin{tabular}{| l | c | c | c | c |}
 \hline			
   emitter [reference] & $\Delta\omega_\mathrm{hom}$ & $\Delta\omega_\mathrm{inhom}$ & $\tau_\mathrm{SD} $ & $R_\mathrm{eq}$ (Hz)\\
    \hline			

   hBN B center~\cite{Fournier23PRB} & 88~MHz & 4~GHz & 30~$\mu$s & 250 \\
   hBN red-emitting SPE~\cite{Spokoyny20} & 5~$\mu$eV & 50~$\mu$eV & 23-510~$\mu$s & 290-6400 \\
   NV center in diamond~\cite{Wolters13} & 13~MHz & 367 GHz & 4.6~$\mu$s & 1160 \\
   Self-assembled QDs~\cite{Kuhlmann13} & 0.93~$\mu$eV & 1.7~$\mu$eV & 100~$\mu$s & 8000 \\
   CdSe nanocrystals~\cite{Beyler13} & 20~$\mu$eV & 200~$\mu$eV & 1~ms & 150 \\
   Perovskite nanocrystals~\cite{Utzat19} & 2.4-3.4~$\mu$eV & 2~meV & 0.5~ms & 2000 \\
   GaN QDs~\cite{Gao19} & 160 ~MHz & 1.7~meV & 260~ns & 2200 \\
   
 \hline  
 \end{tabular}
\end{center}

~\\
The values of $R_\mathrm{eq}$ are in most case experimentally accessible count rates, in particular since our approach does not require to drive the emitter well below saturation, nor to collect only the zero-phonon-line photons (which, for emitters like NV centers where most photons are emitted in phonon sidebands, would lead to a much smaller signal). 
We note that the condition $R > R_\mathrm{eq}$ is necessary for observing both SD-induced intensity noise and SD-induced skewness (relevant for GRJ model) or absence thereof.

\clearpage

\section{S4. Blinking emitters}

Blinking is well known to also affect the long-time $g^{(2)}$ of  emitters subjected to it. It can be accounted for in our approach by including an intensity modulation of the fluorescence. The latter can be in some cases correlated with spectral diffusion.
Experimentally, blinking can often be characterized separately from spectral diffusion, using a non-resonant laser in photoluminescence, or by resonantly driving the emitter using a broad mode-locked laser, which makes the fluorescence intensity insensitive to the emitter spectral position. The associated $g^{(2)}(\tau)$ can then be measured using photon correlations.

\begin{figure}[h]
  \centering
  \includegraphics[width=0.5\linewidth]{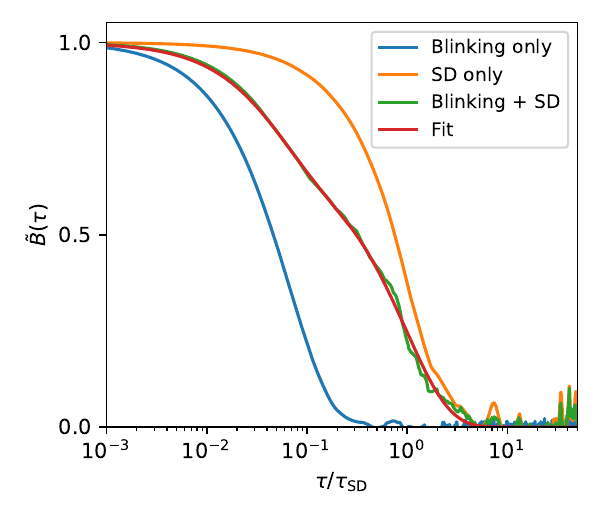}\\
  \caption{Simulated normalized bunching of a blinking emitter (blue curve), of a spectrally diffusing emitter (orange curve) and of an emitter subject to both processes. Red curve: biexponential fit to the simulation.}\label{blinking}
\end{figure}

In this section, we briefly illustrate such a situation, where we have purposely chosen different timescales for blinking and spectral diffusion for clarity (with $\tau_\mathrm{blinking} = \tau_\mathrm{SD}/10)$. Fig.~\ref{blinking} presents the simulated normalized bunching $\tilde{B}(\tau)$ of a blinking emitter (without SD, or under non-resonant excitation -- blue curve), together with $\tilde{B}(\tau)$ for a non-blinking but spectrally diffusing emitter (orange curve) as well as $\tilde{B}(\tau)$ for an emitter experiencing both processes (green curve). In the latter case, the decay is biexponential, exhibiting contributions from both blinking and SD. Note that only the latter contribution would vary with varying laser detuning or power, which is an additional way to experimentally disentangle the two contributions.

\clearpage

\section{S5. Multiple spectral diffusion processes}

\begin{figure}[h]
  \centering
  \includegraphics[width=0.9\linewidth]{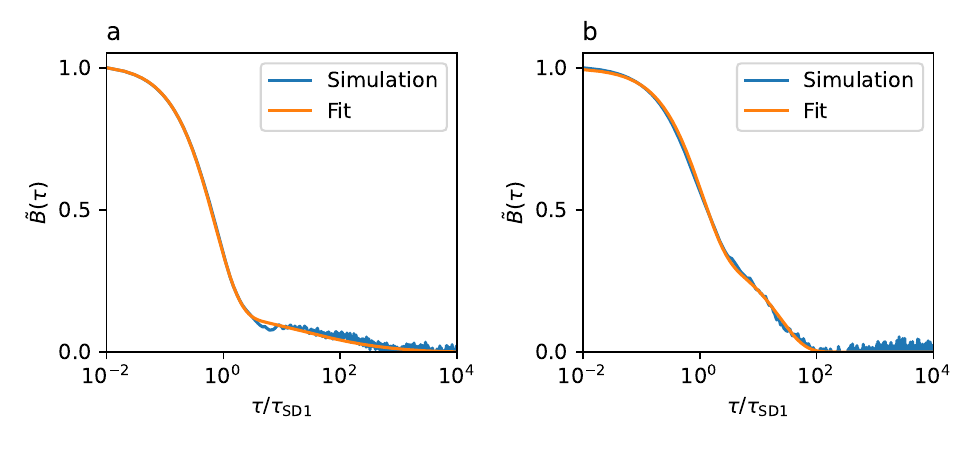}\\
  \caption{Simulated normalized bunching of an emitter subject to (a) a GRJ and a OU process, and (b) two GRJ processes.}\label{multipleSD}
\end{figure}

In this section, we consider an emitter subject to two SD processes with different timescales. Such situation has been encountered in prior work~\cite{Spokoyny20, Abbarchi12}. We performed simulations by including two contributions to the SD -- a combination of continuous and discrete diffusion, and a combination of two discrete processes. Fig.~\ref{multipleSD} shows the simulation results for the bunching function $\tilde{B}(\tau)$, where two decays can be observed in both cases. In the first case, both an exponential and a sub-exponential decay can be observed, while in the second case the decay is biexponential. Although we did not perform a full theoretical analysis, we observe that $\tilde{B}(\tau)$ can be fitted by a sum of both contributions, using Eqs.~8-9 for the OU process, and exponential decays for GJR processes.

\end{document}